\documentclass[twocolumn]{layout/tudelft-aiaa}
\usepackage[T1]{fontenc}

\usepackage{wrapfig}
\usepackage{multirow,tabularx}

\usepackage{amssymb}

\title{StyleDEM: a Versatile Model for Authoring Terrains}

\author{Simon Perche, Adrien Peytavie, Bedrich Benes, Eric Galin \& Eric Gu\'{e}rin}

\usepackage{xcolor}

\begin{document}

\definecolor{turquoise}{cmyk}{0.5,0,0.1,0.2}
\definecolor{purple}{rgb}{0.7,0,0.6}
\definecolor{darkgreen}{rgb}{0.0, 0.5, 0.0}
\definecolor{darkred}{rgb}{0.8, 0.1, 0.0}
\definecolor{darkblue}{rgb}{0.0, 0.2, 0.4}
\definecolor{greenblue}{rgb}{0.0, 0.5, 0.5}
\definecolor{orange}{rgb}{1.0,0.7,0.0}
\definecolor{green}{rgb}{0.0,0.7,0.0}
\definecolor{black}{rgb}{0.0,0.0,0.0}
\definecolor{blue}{rgb}{0.0,0.2,0.4}

\newcommand{\hide}[1]{{}}
\newcommand{\del}[1]{\st{#1}}
\newcommand{\gmp}[1]{\varphi_{\text{\tiny #1}}}

\newcommand{\BB}[1]{{{ \color{green}[BB: #1]}}}
\newcommand{\bb}[1]{{{ \color{green}[BB: #1]}}}

\newcommand{\gon}{}
\newcommand{\goff}{}

\newcommand{\bbon}{\color{green}}
\newcommand{\bboff}{\color{black}}

\newcommand{\bgon}{\color{green}}
\newcommand{\bgoff}{\color{black}}

\newcommand{\eon}{\color{purple}}
\newcommand{\eoff}{\color{black}}

\newcommand{\aon}{\color{orange}}
\newcommand{\aoff}{\color{black}}

\newcommand{\sson}{\color{darkgreen}}
\newcommand{\ssoff}{\color{black}}

\newcommand{\good}{\color{darkblue}}
\newcommand{\average}{\color{orange}}
\newcommand{\bad}{\color{red}}

\newcommand{\derived}[1]{{\color{orange}#1}}
\newcommand{\stated}[1]{{\color{green}#1}}

\font\tenbb=msbm10 \font\sevenbb=msbm7
\font\fivebb=msbm5 \newfam\bbfam
\textfont\bbfam=\tenbb
\scriptfont\bbfam=\sevenbb
\scriptscriptfont\bbfam=\fivebb
\def\bbb{\fam\bbfam\tenbb}
\newcommand{\R}{{\bbb R}}

\newcommand{\rain}{{\omega}}

\newcommand{\ea}{{\mathbf a}}
\newcommand{\eb}{{\mathbf b}}
\newcommand{\ep}{{\mathbf p}}
\newcommand{\eq}{{\mathbf q}}
\newcommand{\ek}{{\mathbf k}}

\newcommand{\ey}{{\mathbf y}}
\newcommand{\ez}{{\mathbf z}}
\newcommand{\es}{{\mathbf s}}
\newcommand{\ec}{{\mathbf c}}
\newcommand{\ex}{{\mathbf x}}

\newcommand{\eu}{{\mathbf u}}
\newcommand{\ev}{{\mathbf v}}
\newcommand{\ew}{{\mathbf w}}

\newcommand{\cV}{{\mathcal V}}
\newcommand{\cP}{{\mathcal P}}
\newcommand{\cC}{{\mathcal C}}
\newcommand{\cS}{{\mathcal S}}
\newcommand{\cE}{{\mathcal E}}
\newcommand{\cR}{{\mathcal R}}
\newcommand{\cG}{{\mathcal G}}
\newcommand{\cA}{{\mathcal A}}
\newcommand{\cN}{{\mathcal N}}
\newcommand{\cL}{{\mathcal L}}
\newcommand{\cT}{{\mathcal T}}
\newcommand{\cB}{{\mathcal B}}
\newcommand{\cQ}{{\mathcal Q}}
\newcommand{\cD}{{\mathcal D}}
\newcommand{\cH}{{\mathcal H}}
\newcommand{\cU}{{\mathcal U}}
\newcommand{\cW}{{\mathcal W}}
\newcommand{\Es}{\mathbb{E}}

\newcommand{\bA}{\mathbf{A}}
\newcommand{\bB}{\mathbf{B}}
\newcommand{\bX}{\mathbf{X}}
\newcommand{\bY}{\mathbf{Y}}
\newcommand{\bZ}{\mathbf{Z}}

\def\tvi{\vrule height 12pt depth 5pt width 0pt}
\def\tvs{\vrule height 10pt depth 1pt width 0pt}

\newcommand{\Fig}{Figure}
\newcommand{\Sec}{Section}

\renewcommand{\um}{{\mathrm {m}}} 
\renewcommand{\us}{{\mathrm {s}}} 
\newcommand{\ums}{{\mathrm {ms}}} 
\newcommand{\ukm}{{\mathrm {km}}}
\newcommand{\ucm}{{\mathrm {cm}}}
\newcommand{\uGb}{{\mathrm {Gb}}}
\newcommand{\uGHz}{{\mathrm {GHz}}}

\newcommand{\hold}{\color{orange}}
\newcommand{\warn}{\color{darkred}}

\newcommand{\qa}[2]{#2}
\newcommand{\qatry}[2]{#2}
\newcommand{\qaok}[2]{#2}
\newcommand{\qacap}[1]{{#1}}
\newcommand{\qacaptry}[1]{{#1}}
\newcommand{\qacapok}[1]{{#1}}

\marginparwidth=10mm
\newcommand{\cqa}[2]{{\color{purple}\textbf{#1}}: {\color{purple}#2}}
\newcommand{\rev}[1]{{{\color{blue}#1}}}

\newcommand{\ie}{\textit{i.e.}}
\newcommand{\eg}{\textit{e.g.}}

\newcommand{\vol}{{$3$D}\phantom{ }}

\newcommand{\dem}{{\color{black}\cT}}
\newcommand{\tildedem}{{\color{black}\widetilde{\cT}}}
\newcommand{\sketch}{{\color{black}\cS}}
\newcommand{\latent}{{\color{black}\cW}}
\newcommand{\latentplus}{{\color{black}\cW^+}}
\newcommand{\patch}{{\color{black}\cP}}
\newcommand{\latentvectoru}{\eu}
\newcommand{\latentvectorv}{\ev}
\newcommand{\latentvector}{\ew}

\newcommand{\etal}{\emph{et al.}}

\newcommand{\styledem}{{\color{black}StyleDEM}}
\newcommand{\stylegan}{{\color{black}StyleGAN}}
\newcommand{\terrain}{{\color{black}terrain}}
\newcommand{\terrains}{{\color{black}terrains}}
\newcommand{\super}{{\color{black}super-resolution}}

\newcommand{\redtxt}[1]{\textcolor{red}{#1}}
\newcommand{\citationneeded}[1][]{\textsuperscript{\color{blue} [citation needed: #1]}}

\newcommand{\compress}{\vspace{-5.0mm}}


 \AlwaysPagewidth{
\maketitle

\begin{abstract}
	\noindent Many terrain modelling methods have been proposed for the past decades, providing efficient and often interactive authoring tools. However, they generally do not include any notion of style, which is a critical aspect for designers in the entertainment industry.
We introduce \styledem, a new generative adversarial network method for \terrain\ synthesis and authoring, with a versatile toolbox of authoring methods with style. This method starts from an input sketch or an existing \terrain. \gon It outputs a \terrain\ with features that can be \goff authored using interactive brushes and enhanced with additional tools such as style \gon manipulation or super-resolution. \goff The strength of our approach resides in the versatility and interoperability of the toolbox.
\newline
\newline
\end{abstract}

 }


\section{Introduction}

Realistic and controllable \terrain\ models are necessary for creating virtual worlds. The existing approaches encompass a variety of methods, including procedural generation, physically-based erosion simulations, and example-based synthesis. Existing methods provide a varying level of control, allowing the user to generate, edit, or modify synthetic \terrains. However, they generally do not include any notion of style, which is a critical demand from designers in the entertainment industry. 

We define style intuitively as related to the user perception of the overall quality and common properties of a \terrain\ or a specific category with similar elevation and landmark characteristics. A key observation is that style has a tremendous impact on the perception of \terrains: a highly irregular mountainous landscape from the Karakoram looks radically different from smooth hills in the Appalachians. Style is not only present at a large scale like in the orometry-based method proposed in\,\cite{Argudo2019}, but it affects all ranges of scales, from large-scale geographic structures of hundreds of kilometres to landforms of a few meters. This diversity in style is the consequence of complex natural processes acting over varying temporal and spatial scales, including tectonics, stratification, aeolian and hydraulic erosion. 
Modelling such complex phenomena is intricate and comes at the price of computationally intensive, involved, and hard to understand and control simulations. Most erosion algorithms focus on restricted temporal and spatial scales. In effect, they only account for limited phenomena and simple materials, thus allowing only for the simulation of a limited range of landforms. 

\begin{figure}[h!t]
 \centering
 \includegraphics[width=84mm]{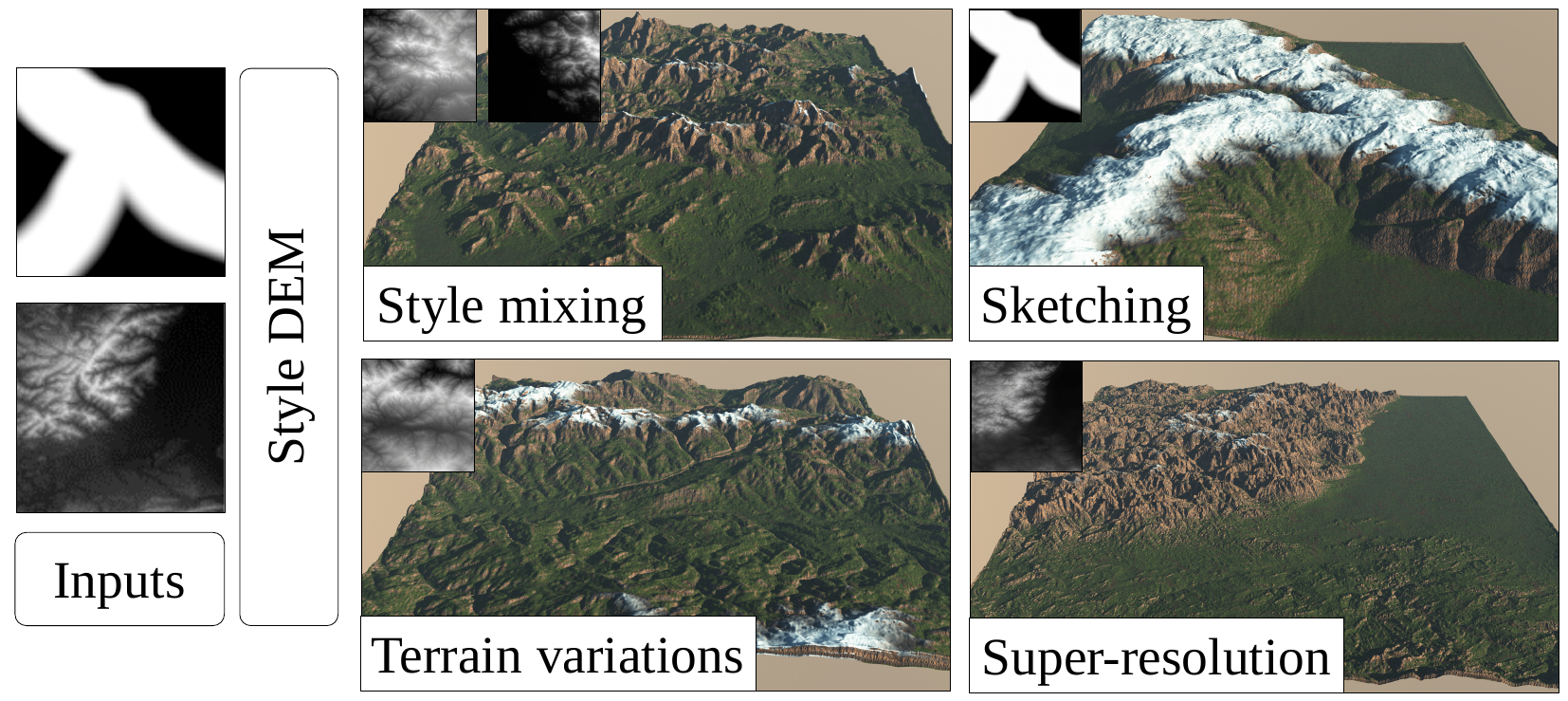}
 \caption{\styledem\ is a deep-learning model that offers versatile authoring possibilities. A range of tools is offered to users so they can easily author high-quality terrains.}
 \label{image:teaser}
\end{figure}

Another crucial observation is that artists prefer an interactive editing process when authoring \terrains\ and therefore favour techniques that allow user control. While methods exist that allow for interactive modifications of landform features~\cite{Gain2009} or local style transfer for virtual worlds~\cite{Emilien15ToG}, little effort has been dedicated to combining the concept of style with an interactive edition framework. We address these limitations by proposing a novel approach that presents versatile interactive tools for editing \terrains, including sketching, copy-and-paste sequence, and super-resolution (adding details) while providing ways for the designer to define or impose a given style at different levels (\Fig~\ref{image:teaser}).

More precisely, our contributions are: 
1) a novel \terrain\ model employing a \stylegan\ architecture that describes and generates high-quality digital elevation models; 
2) a variety of tools adapted to the model that are central in \terrain\ authoring; 
3) the introduction of \terrain\ style at different scales, which allows for rich context-aware content generation.


\section{Related work}

Terrain generation methods in Computer Graphics can be classified as procedural generation, physically-based 
(erosion) simulation and synthesis from exemplars, which includes deep-learning algorithms.
Given the identified goal of control, style, and synthesis from real digital elevation models, we briefly review the first two categories and focus on authoring frameworks that evidence interactivity. We refer the reader to \,\cite{Galin2019} for a complete overview of \terrain\ generation techniques.

\paragraph*{Procedural terrain modelling} relies on procedural noise, often combined with river network carving, to algorithmically reproduce
the self-similarity across scales. Most techniques rely on a globally defined sum of noises\,\cite{Musgrave1989} or assembly of 
procedurally defined and compactly supported primitives\,\cite{Genevaux2015}. They are generally computationally efficient and lend themselves to parallel implementation on graphics hardware.
Authoring control typically takes the form of applying noise with circular brushes\,\cite{Carpentier2009} or matching curve and point constraints
using warping\,\cite{Gain2009},
or using diffusion curves\,\cite{Hnaidi2010}.
Recently, \cite{Guerin2022} introduced gradient-domain editing for \terrain\ modelling, demonstrating its effectiveness 
for seamless blending of 
patches, copy-and-paste operations, and generation from control feature points and curves.
Unfortunately, these approaches do not consider style and require careful and tedious editing and parameter tuning to obtain realistic landmarks. 

\paragraph*{Erosion simulation} methods can be broadly classified into surface erosion algorithms and tectonic-based simulations.
Surface erosion\,\cite{Musgrave1989} simulate material detachment, transport, and deposition, possibly considering  the strata of the bedrock \cite{Roudier1993}, and enhance relief with sedimentary valleys and small-scale erosion landmarks such as gorges and ravines. 
These early approaches were improved in several ways by computing the 
acceleration or deceleration of the fluid to erode the bedrock or deposit sediments~\cite{Neidhold2005}, 
combining a shallow water simulation with hydraulic erosion~\cite{Benes2007}, or Smoothed Particle Hydrodynamics~\cite{Kristof2009}. 
Those methods generate convincing small-scale erosion effects as long as the initial input \terrain\ is sufficiently realistic and supplies large-scale landmarks. 
Our method addresses this issue by implicitly providing erosion-generated features learned from real-world examples via neural transfer.
 
Tectonic simulations, in contrast, attempt to reproduce large-scale erosion by taking into account the uplift of the bedrock produced by relative movement of the tectonic plates, balanced by different types of erosion, most often the Stream Power erosion~\cite{Cordonnier2018}. 
Contrary to surface erosion, tectonic simulations produce realistic large-scale mountain ranges without needing an initial elevation. 
However, both methods are difficult to control and computationally intensive, which deters them from interactive authoring. 
In contrast, our method implicitly addresses the control via deep learning transfer from real-world features.

\paragraph*{Example-based methods} tackle the \terrain\ realism by combining patches extracted from real-world digital elevation models.
Their control is achieved by structure-sensitive warping to match sketched silhouettes\,\cite{Tasse2014}, 
use of Conditional Generative Adversarial Networks to learn the correspondence
between \terrains\ and the sketch maps corollaries containing ridge
and river lines and feature points \cite{Guerin2017}. Parallel texture-based 
synthesis~\cite{Gain2015} 
modifies the matching process to support style painting, region-based copy-and-paste, and curve and point manipulators. 
The assembly of \terrain\ patches, even locally geomorphologically correct, is insufficient for generating globally consistent landscapes.
Sparse modelling is another efficient way to generate high-resolution \terrains\ from sketches\,\cite{Guerin2016} guided by exemplars.
Recently, Scott \etal \cite{Scott2021} proposed a breaching algorithm interlaced in multi-resolution example-based \terrain\ 
synthesis to improve hydrological consistency. 
In contrast, our method encodes hydrological consistency at different scales in the latent space.

\paragraph*{Deep-learning algorithms} are a specific case of example-based approaches. 
Zhang \etal~\cite{zhang_authoring_2022} used a modified version of a GAN with low-resolution maps, global style information, and local style maps as input, and a discriminator capable of classifying different types of \terrains. The generator is based on UNet architecture, and patch-based discriminators allow for the local control of style. 
Another approach, specialised in style embedding, was proposed in \cite{Zhao2019}. By using a cGAN to insert the embed theme into the \terrain, the method can amplify an input low-resolution \terrain\ into a high resolution with style variation. While producing compelling, high-quality results in terms of style transfer, this approach does not provide any edition tools. 
Recently, \cite{naik_deep_2022} trained a Variational Auto Encoder \cite{kingma2013auto} combined with a GAN to generate a heightmap from a low-resolution map coupled to a sketch. While providing authoring tools such as sketching and \terrain\ interpolation, results lack details. 
We propose a novel method that addresses the blind spot of previous approaches by developing a framework for simultaneous style manipulation and \terrain\ editing.


\section{Model}

We introduce \styledem, a deep neural model that is based on the \stylegan\ architecture~\cite{karras_stylegan_2019} applied to Digital Elevation Models (DEMs), which are \terrains\ represented as discrete heightfields denoted as~$\dem$. \stylegan\ takes a large set of input images and encodes their style into its latent space. Inspired by this approach, we use latent space as the primary way to represent and manipulate digital \terrains. 

\begin{figure}[h!t]
 \centering
 \includegraphics[width=84mm]{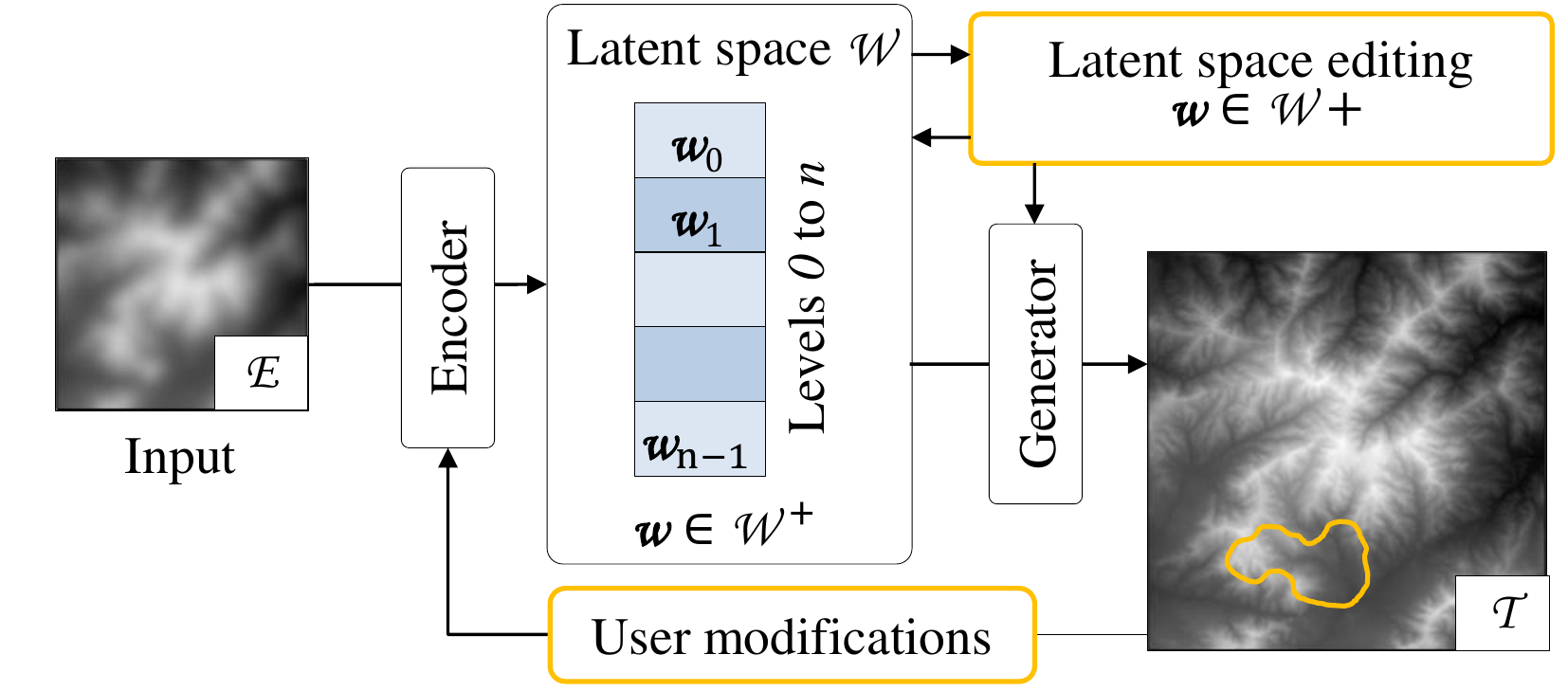}
 \caption{The input heightfield $\cE$ (a high or a low-resolution \terrain, or even a sketch) is injected into the encoding process to produce a latent vector representation in $\latentplus$.
 The generator uses one or multiple $\latentplus$ vectors to synthesise a new \terrain\ $\dem$. The user may modify the latent space to control the generation process, or directly edit $\dem$ and inject it again in the generator. }
 \label{sketch:overview}

\end{figure}

 Our work consists of two main parts depicted in Figure~\ref{sketch:overview}. The \stylegan\ itself is a \emph{generator} trained on a carefully selected and designed dataset of DEMs. The generator synthesises a high-resolution \terrain\ $\dem$ from its latent representation $\latent$. 
Conversely, an inverter called \emph{encoder} takes a latent vector corresponding to the \terrain\ as its input and generates the output DEM. The user can interact at multiple levels of this pipeline, from user inputs to direct latent space modification. At each step, the output is directly rendered for preview or, when the result is satisfactory, streamed to a rendering pipeline. \goff

 \begin{figure}[ht!]
	\centering
	\includegraphics[width=84mm]{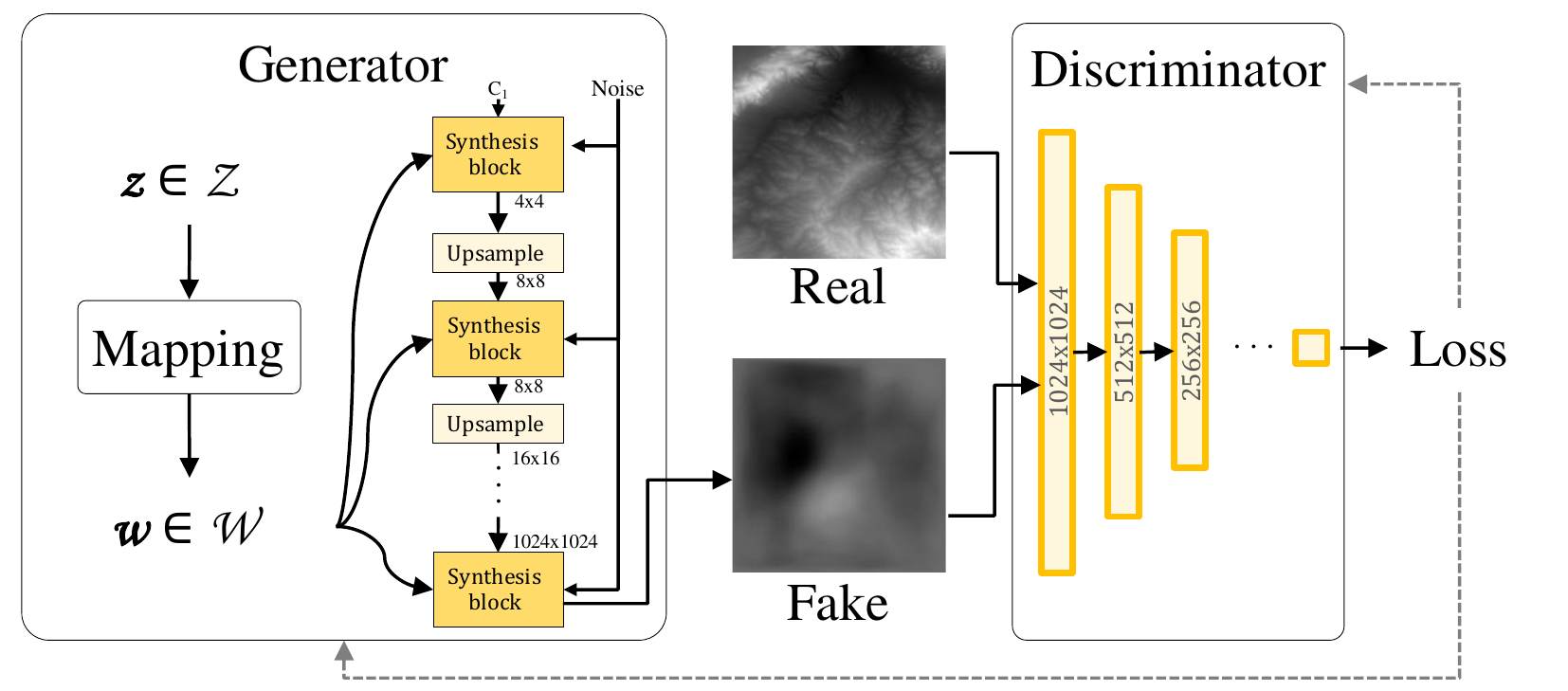}
	\caption{StyleGAN architecture. A vector from the latent space $\mathcal{Z}$ is projected into a second latent space $\latentvector \in \latent$ by using a mapping network that disentangles latent directions. This $\latentvector$ vector is then fed to the generator at various level of resolution, and the output is sent to a discriminator, which seeks to separate real and fake image. The generator and the discriminator are trained together in a zero-sum game. The results are utilised during the backward propagation phase. }
	\label{image:sg2_architecture}
\compress
\end{figure}

\paragraph*{Generator}
		The generation step incrementally builds on the \emph{Generative Adversarial Network} (GAN) architecture introduced in \cite{goodfellow_generative_2014}.
 The generator denoted as $\mathcal{G}$ creates a high-resolution \terrain. 
 During the training process, $\mathcal{G}$ is coupled with another neural network, called the discriminator $\mathcal{D}$ that attempts to detect whether the generated \terrain\ is real or not. Therefore, the generator and the discriminator compete against each other: the generator tries to fool the discriminator by generating images resembling the ones found in the training database, whereas the discriminator tries to distinguish images generated by the generator from those in the training dataset. 
 This architecture allows the network to be
trained in an unsupervised way and does not require explicit specification of a loss function.	 
 
The introduction of the \stylegan\ architecture \cite{karras_stylegan_2019} significantly improved the GAN model and demonstrated its performance by being the first to create photorealistic images while maintaining user control by taking advantage of its latent space $\latent$.
Here we extend the scope of \stylegan\ to digital \terrains. 
One particularity is the progressive growth of the output image performed by generators at increasing resolutions that take the previous layers as input, driven by the latent representation in $\latent$. 
Starting at a reduced initial low resolution (in general $4\times{4}$), every step increases the resolution of the filters by a factor of two, and the generator employs the latent vector $\latentvector$ to add details. Another layer is specialized to transform these filters into images. \Fig~\ref{image:sg2_architecture} illustrates this particular architecture.

Unfortunately, as demonstrated in \cite{abdal2019image2stylegan}, $\latent$ cannot represent every real image. Therefore, $\latentplus$ vectors were introduced to express the necessary variability across models. 
 The \stylegan\ generator uses the $\latentplus$ as a concatenation of 18 different 512-dimensional $w$ vectors for each layer.
 `gon The vectors in $\latentplus$ control the generation of every layer over the hierarchical process.\goff

\paragraph*{Encoder}
\label{par:inverter}

\begin{figure}[ht!]
	\centering
	\includegraphics[width=84mm]{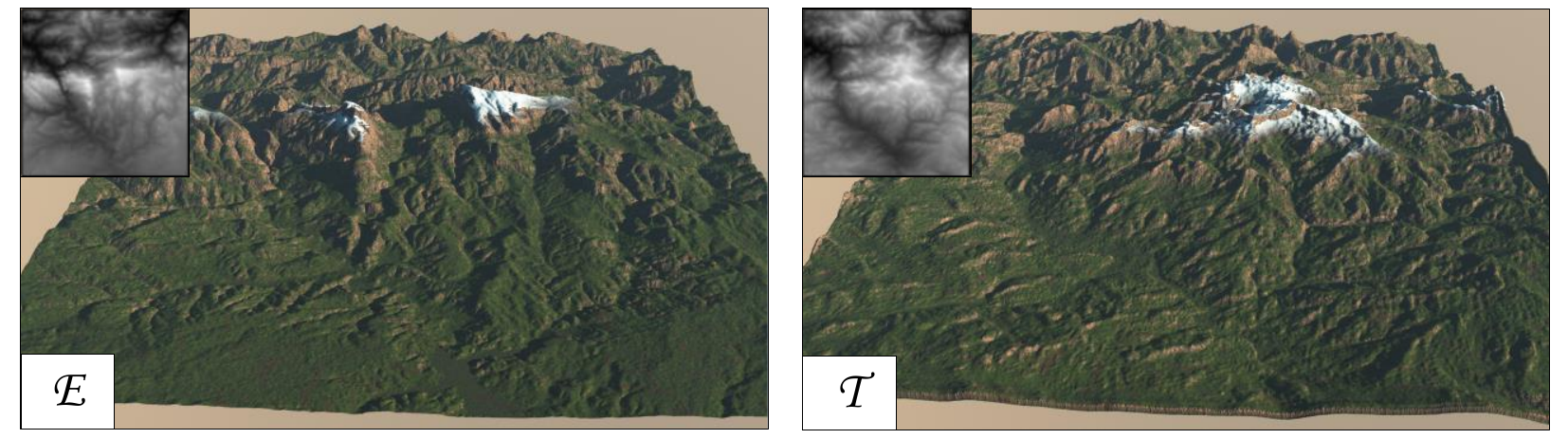}
	\caption{The encoder inverts an input terrain (left) into latent space, before the generator synthesises an approximation (right) using the latent vector representation.}
	\label{image:encoder}
\compress
\end{figure}

 The \stylegan\ architecture has a great synthesis power but lacks a so-called \emph{inversion} process: what is the representation of an existing \terrain\ in the latent space? 
 Two main categories of inversion methods exist. Optimisation-based approaches compute the loss between the generated and target \terrain. Starting from an initial random vector in $\latentplus$ and using an optimiser and back-propagation, the system performs an optimisation over $\latentplus$. Despite its high-quality results, this method is computationally intensive, up to several minutes per image, which is prohibitive for an interactive application. 
We prefer the encoder-based approach for inverting a \terrain, which implements another neural network, denoted as $\mathcal{I}$, encoding an image into a $\latentplus$ vector. This method necessitates preliminary training and thus operates on pairs of \terrain\ model and latent representations. 
 We refer the reader to a review of GAN inversion from \cite{xia_gan_2022} for more details. \goff
 The encoder option lends itself to the interactive generation pipeline because of its efficiency. Moreover, it allows for generalisation, because it can invert various inputs, including high or low-resolution DEMs or sketches. This property is essential in our model and allows for a range of use cases adapted to authoring.  
 We use the pSp (pixel2style2pixel) architecture proposed in \cite{richardson_encoding_2021} that allows the \stylegan\ to produce an output image based on an input image using the latent intermediate representation (See figure~\ref{image:encoder}). In our experiments, we used L2 and LPIPS losses for the encoder that provided the best results.	 
 In the remainder of the paper, $\latentvector$ denotes a vector from $\latentplus$.


\section{Terrain authoring with style}
\label{sec:authoring_style}

We developed and studied a variety of authoring tools that benefit from the latent space representation and the generalisation possibilities of the encoder. Since the information given to the encoder uses the same format as the generator output, it intrinsically allows the interactive edition of the output and iterating through the process. 

\subsection{Versatility}
\label{subsec:handmade_drawing}
We trained the encoder (\Sec~\ref{subsec:data}) using DEMs as input and $\latentvector \in \latentplus$ latent vector as output. Every vector $\latentvector$ encodes a high-resolution topography in the high-dimensional latent space of the generator. An immediate consequence is that the output of \styledem\ inherently retains consistent geomorphological properties. 
We exploit the inference capabilities of the encoder and feed it with new inputs that are different from those used during the training phase. This generalisation is beneficial for our approach as it allows the user to sketch low-resolution maps and edit existing DEMs while keeping consistency and generating necessary details. 
 
\begin{figure}[ht!]
 \centering
 \includegraphics[width=84mm]{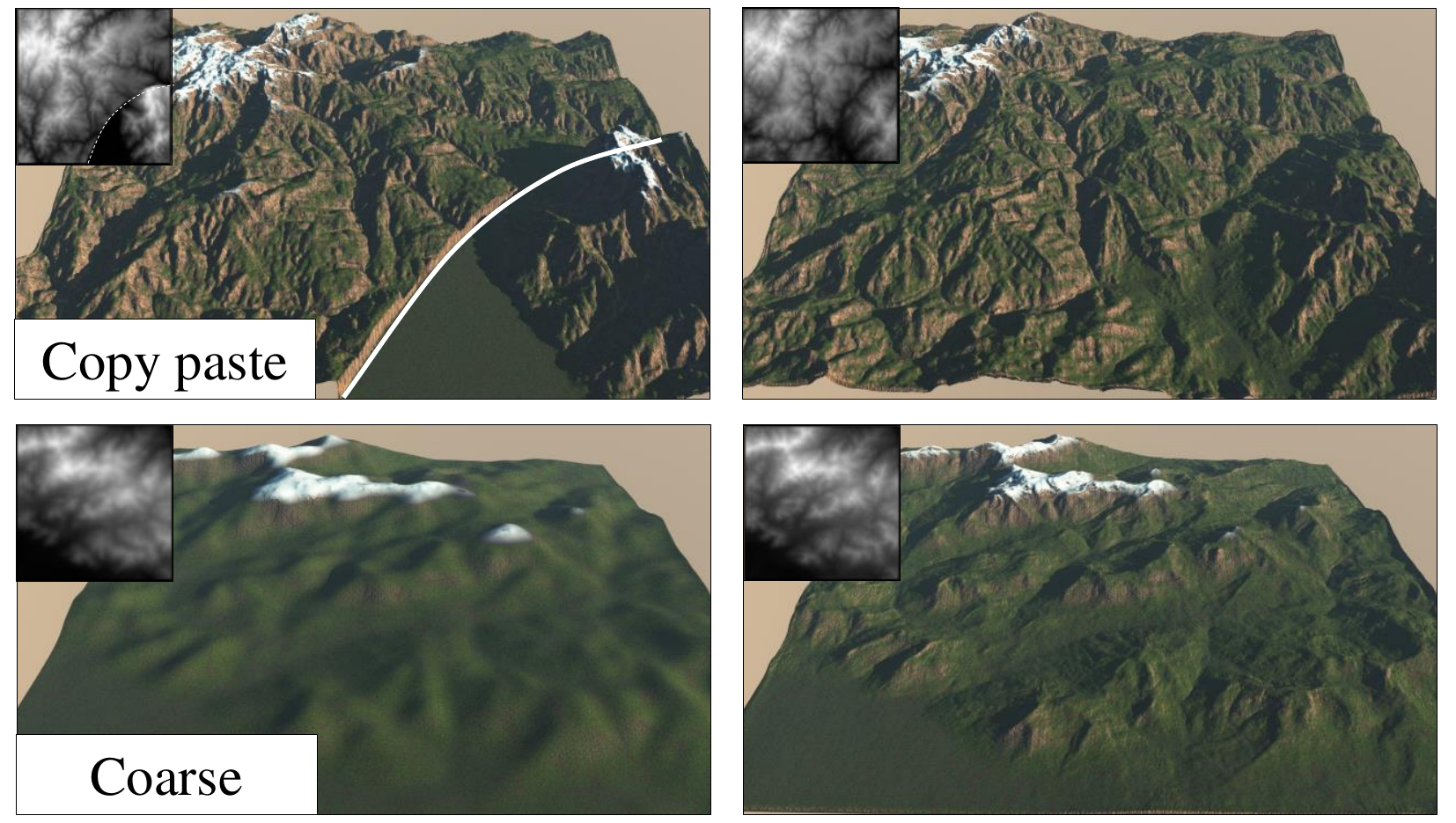}
 \caption{Different categories of input \terrains\ (left) and the corresponding topographies produced by the full \styledem\ encoder and generator process (right): synthesised models exhibit more small-scale details and landforms while maintaining a global consistency.}
 \label{image:versatility}

\end{figure}

\Fig~\ref{image:versatility} demonstrates that \styledem\ can be employed in a variety of use cases: a DEM from a real \terrain, a copy-and-paste editing, a user-defined sketch, and a low-resolution DEM. While a common strategy consists in utilizing an existing \terrain\ before progressively modifying it, our method also allows using any of the different input data types designed by an artist while keeping the same format. 
In previous work, copy-and-paste operations either introduce seams (unless performed in the gradient domain as proposed in \cite{Guerin2022}) or require the use of a blending region to smoothly interpolate the elevations of argument \terrain\ patches\,\cite{Genevaux2015} which often results in inconsistent characteristics or cross fade of styles. In contrast, \styledem\ produces elevations that conform to the essence of the argument \terrain\ patches. 
Moreover, artists often start with sketches depicting prominent features, such as mountain landmarks. In this configuration, 
the networks naturally synthesises \terrains\ that follow the user design approach. 
By resembling the input sketch at every step, it provides consistent and realistic geomorphologic features. Eventually, when fed with a low-resolution model, the networks automatically synthesise a \terrain\ approximating the input and enhanced with details. While in spirit similar to augmenting a \terrain\ with small-scale features using sparse modelling\,\cite{Guerin2016} or procedural noise, the output preserves the style encoded in the latent space. 

 \begin{figure}[htb]
 \centering
 \includegraphics[width=84mm]{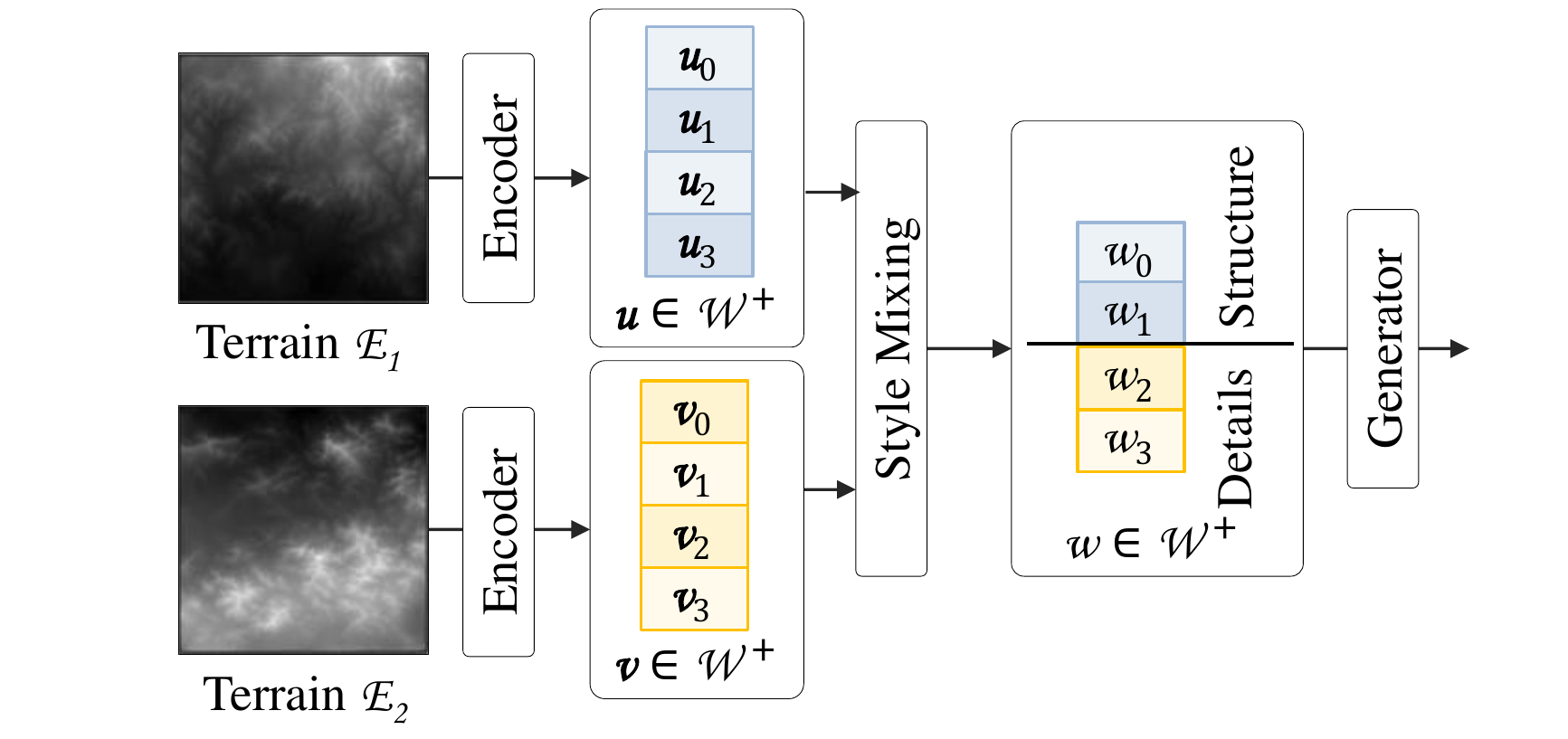}
 \caption{The style mixing proceeds as follows: two latent vectors $\eu$ and $\ev$ are computed using the encoder and then combined into $\ew$ to be finally fed to the generator. The upper part $\eu_0$ to $\eu_{i-1}$ is combined to $\ev_i$ to $\ev_{n-1}$.}
 \label{fig:style_mixing_schema}

\end{figure}

\subsection{Style mixing}
 \label{subsec:style_mixing}
		
Inherited from the \stylegan\ architecture, the generator is structured in $18$ layers controlled by the latent vector. While the same latent vector controls the different layers during the training phase, we can also use different ones, which is the purpose of extending the latent space to $18\times 512$ elements. 
This is particularly true when using the encoder that produces an extended latent vector $\latentvector \in \latentplus$ to increase the expressivity (see \Sec~\ref{par:inverter}). This delivers style mixing capacities: the large-scale structure and landmarks of one \terrain\ can be mixed with the details of another one. The structure is represented in the upper layers of the latent vector, whereas details are connected to the lower ones. 

\begin{figure}[tb]
	\centering
	\includegraphics[width=84mm]{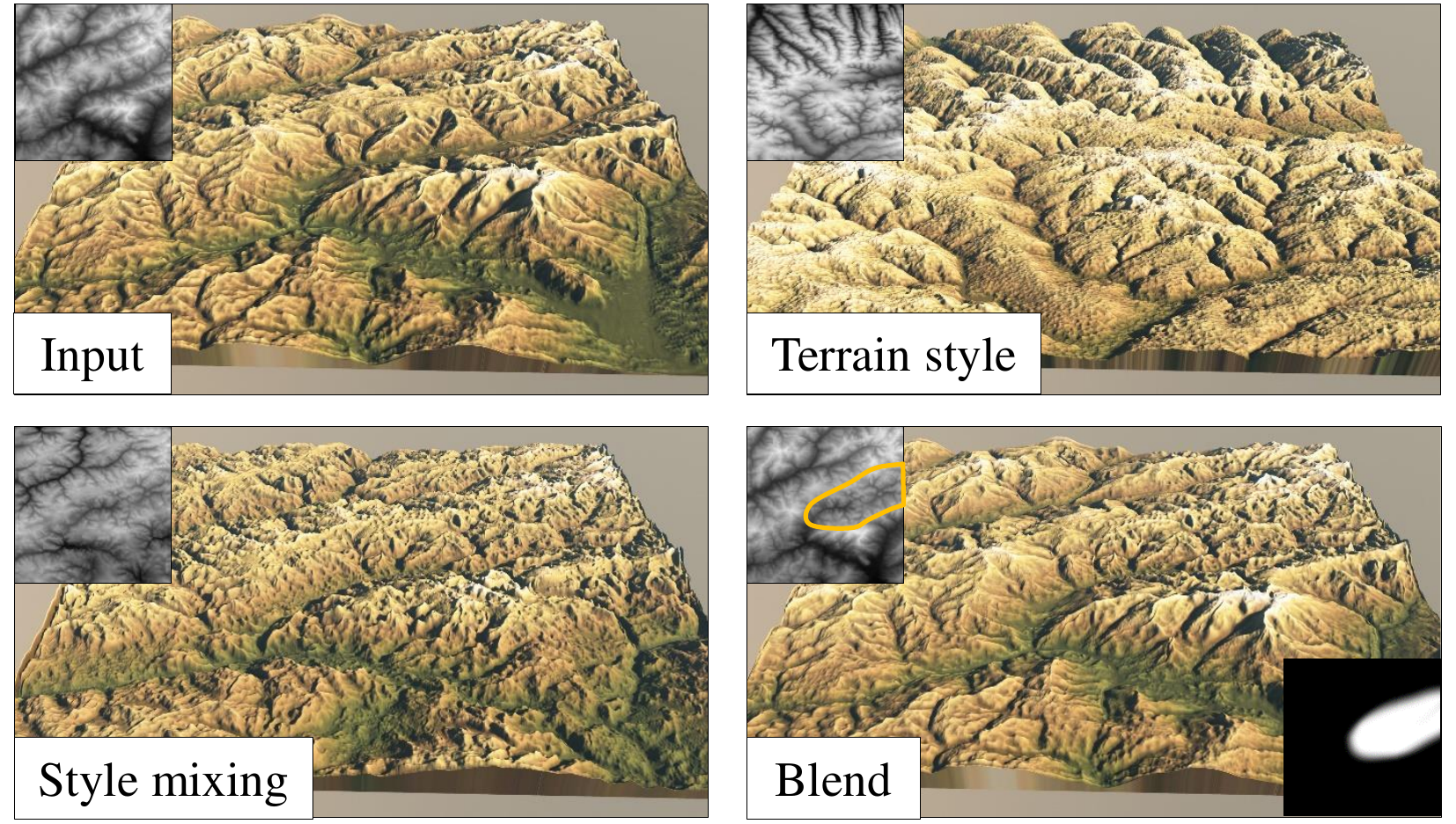}
	\caption{To modify the style over a region $\Omega$ of an input \terrain\ $\dem$, we first apply style mixing over the entire domain to obtain a new terrain $\tildedem$, and then blend arguments with a mask derived from $\Omega$.}
	\label{fig:stylebrush} 

	\end{figure}
	
\begin{figure*}[h!tb]
	\centering
	\includegraphics[width=\linewidth]{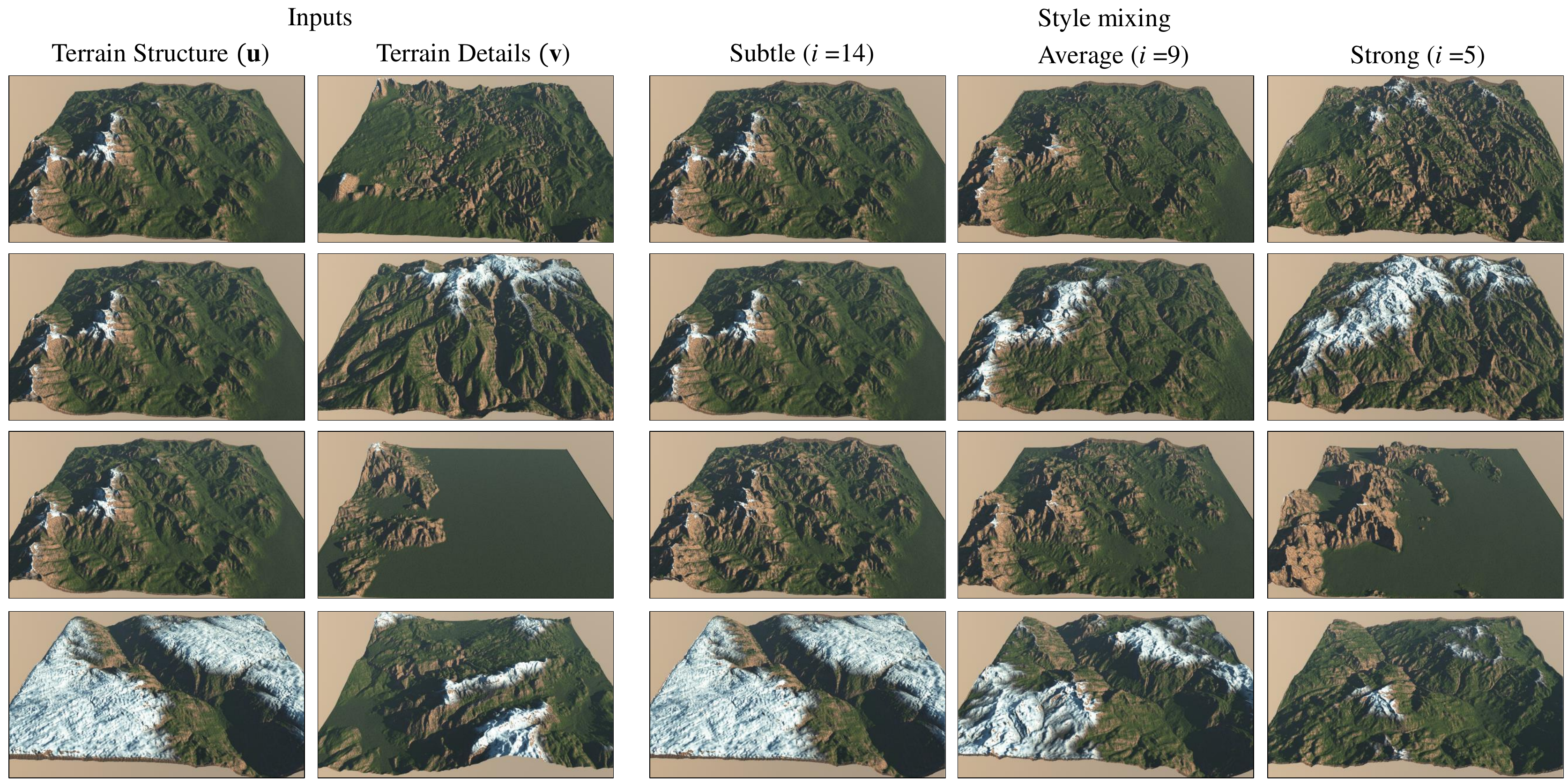} 
	\caption{Influence of the $i$-th level when using style mixing operation. The user can adjust 
   the style, from a subtle effects ($i=14$) to substantial style enforcing ($i=5$). The first three rows share the same input to illustrate the impact of different style mixing. }
	\label{fig:stylemixing-result}
\end{figure*}

We designed a tool that combines the global structure layers from a given \terrain\ with the remaining layers of another \terrain. Thus, we build an extended latent vector by merging two different parts of latent vectors (see \Fig~\ref{fig:style_mixing_schema}). As shown in \cite{karras_stylegan_2019}, the network's first layers (low-resolution) contain large-scale features, and the scale of the features decreases with the successive layers. This is a consequence of the growth of characteristics of the generator. Since small resolutions are generated in the first layers and then upsampled, only large features could be expressed. Therefore, the latent vector $\latentvector$ provided at a given resolution controls the style at this corresponding scale.
For \terrains, it corresponds to decreasing spatial features from mountains and valleys that define the global geomorphology to small-scale details erosion landmarks.
This inherent style embedding of the model allows quick prototyping of \terrains\ by changing details according to an input style, potentially resulting in a completely different visual perception of the initial \terrain.
The user selects the number of layers needed to apply the desired effect on the \terrain, which allows for a balance between global and local control.
\Fig~\ref{fig:stylemixing-result} shows the impact of the number of layers. The generator is fast enough to perform those operations interactively.

One limitation of the \styledem\ and \stylegan\ architectures is the impossibility of mixing styles of the same levels in the same \terrain\ at different locations. 
To overcome this limitation, we allow the user to generate two different stylised \terrains\ and blend them directly in the altitude domain using an alpha mask defined by a brush (see \Fig~\ref{fig:stylebrush}).

\subsection{Interpolation}
\begin{figure*}[htb]
	\centering
	\includegraphics[width=\linewidth]{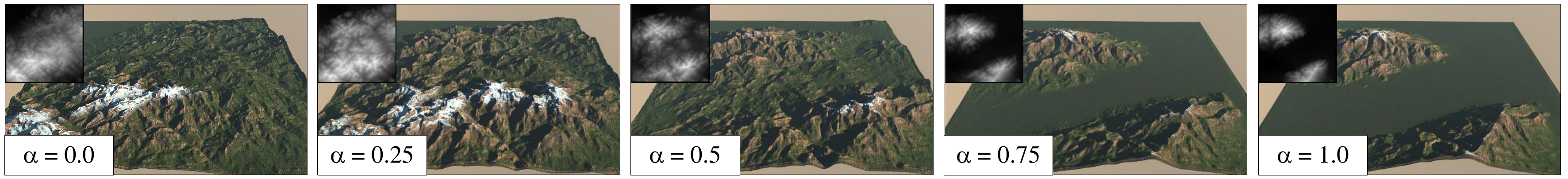}
	\caption{Terrains obtained using the interpolation between two latent space vectors.}
	\label{fig:interpolation}
	\end{figure*}

Another important property of the latent space is the consistency of interpolation as the interpolated latent vector embeds meaningful information about the \terrain. 
Formally, given two latent descriptions $\latentvectoru$ and $\latentvectorv$, we define the interpolated representation as: $\latentvector = (1-\alpha) \; \latentvectoru + \alpha \; \latentvectorv$, which interpolates the latent characteristics. When $\alpha$ varies, the resulting \terrain\ $\dem(\alpha)$ morphs from the first to the second \terrain s as shown in \Fig~\ref{fig:interpolation}. 

Contrary to the direct elevation-based interpolation methods (such as elevation interpolation or optimal transport-based techniques), 
this approach generates intermediate \terrains\ $\dem(\alpha)$ that exhibit plausible geomorphological features due to the coherence within the latent space caused by the disentangled latent space $\latent$ during training.

\subsection{Super-resolution}

Super-resolution, or amplification, is crucial in \terrain\ modelling and refers to simultaneous increasing of the resolution while inserting meaningful and consistent details. This step saves a lot of time for artists as it allows for focusing on the main structures and prominent features of a \terrain. 

We exploit the generalisation capabilities of \styledem\ and particularly the way details can be generated when the input is a low-resolution map. We can treat two different final precisions based on the two \styledem\ models that have been trained: $30$ and $5$ meters resolution.

\begin{wrapfigure}{l}{28mm}
	\centering
	\includegraphics[width=28mm]{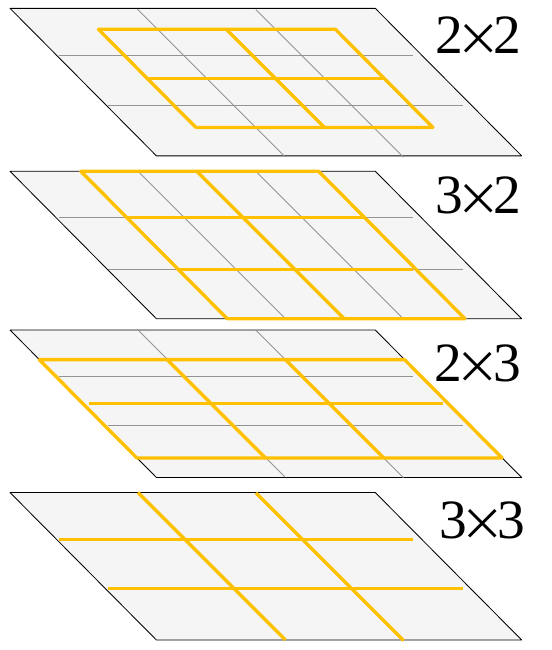}
	\caption{Patch decomposition.}
	\label{sketch:patches}\compress
\end{wrapfigure}
\begin{figure}[h!t]
	\centering
	\includegraphics[width=84mm]{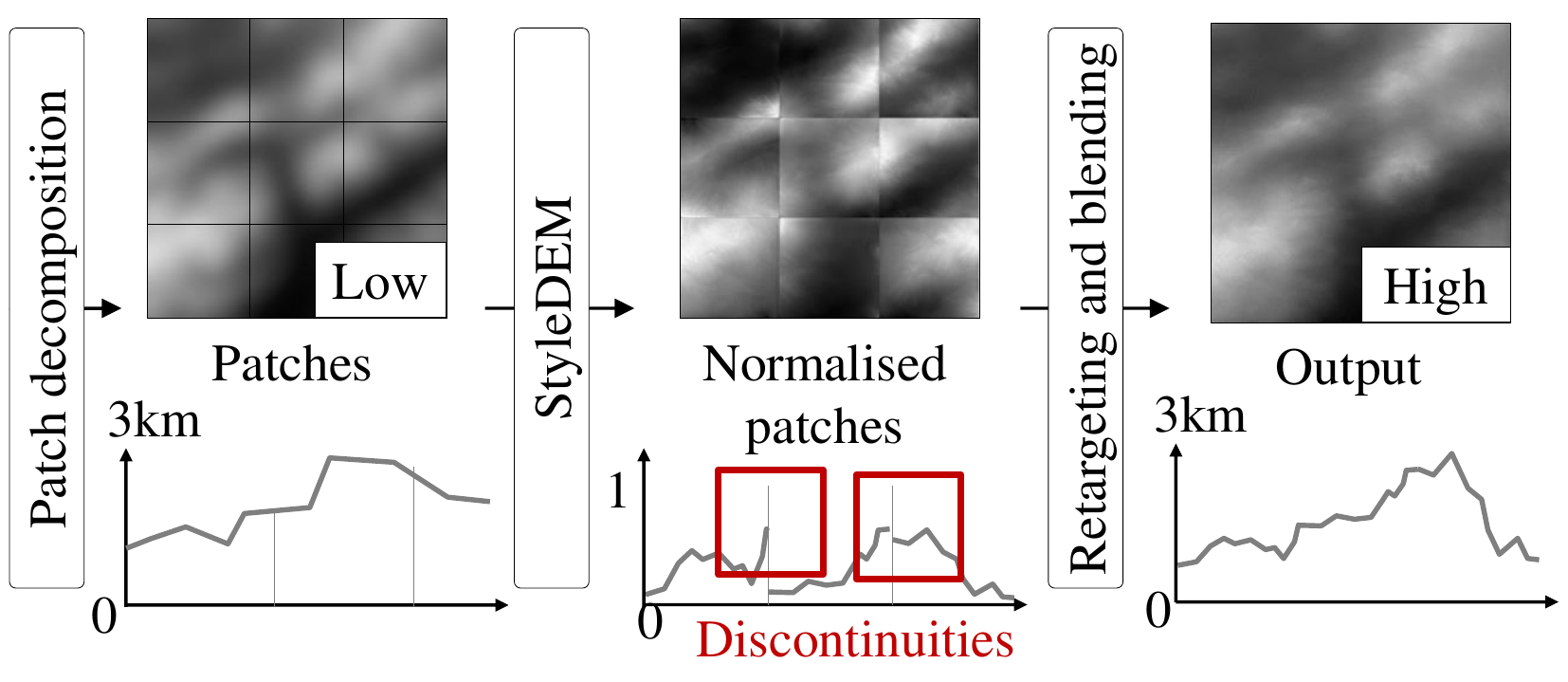}
	\caption{
   The input \terrain\ is divided into $n\times{n}$ patches to adapt the size of the trained model. 
   After processing by the networks, we retarget the heights of the (normalised) high-resolution patches and finally blend patches together.}
	\label{fig:sr_pipeline}
   \end{figure}

Because the model has a fixed resolution of $1,024 \times 1,024$, generated \terrains\ have a size of about $s \approx 30$ and $s \approx 5$ kilometers respectively. To overcome this limitation, we handle larger resolutions and sizes by dividing them into patches of a size $s$ (see \Fig~\ref{fig:sr_pipeline}). 

We decompose input terrains $\dem$ of arbitrary resolution, \ie, larger than the $1,024\times 1,024$ resolution required by the networks, into a grid of $k^2$ patches.
Detailed patches produced by the encoder have a normalised elevation range. 
 
Therefore, we need to retarget each patch elevation to the original patch using a histogram matching, which guarantees recovering of the original elevation range and distribution. 
This strategy still produces discontinuities at the borders of the independently-generated patches. To overcome this limitation, we use a half-size $s/2$ overlapping and blending. We compute intermediate patches covering the boundaries as illustrated in \Fig~\ref{sketch:patches} and combine the three layers of intermediate patches with offset vectors $(s/2)\,\ex$, $(s/2)\,\ey$ and $(s/2)\,(\ex+\ey)$ respectively. The process yields $(2k-1)^2$ patches. We finally stitch them together using the minimum error boundary cut from \cite{Efros2001}.

\begin{figure}[htb]
	\centering
	\includegraphics[width=84mm]{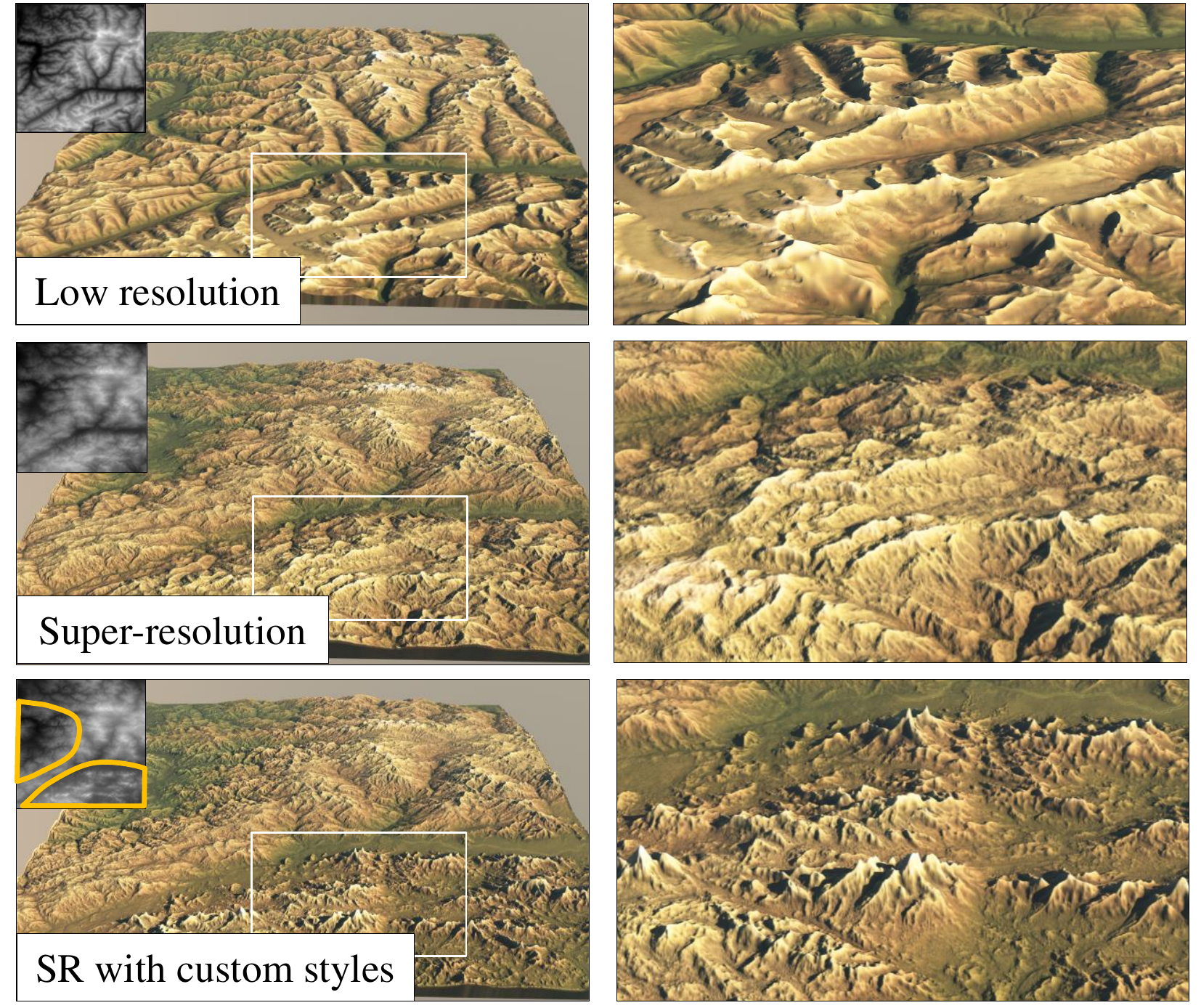}
	\caption{A $90\,\ukm$ terrain of $30\,\um$ precision was obtained with the super-resolution tool (center) from an initial low resolution model (top). We controlled the generation by adding iteratively two different styles (bottom).}
	\label{fig:sr_result}
\end{figure}

\Fig~\ref{fig:sr_result} shows a \terrain\ that was processed by the super-resolution tool which multiplied the resolution by a factor of~$3$.  

A \styledem\ is used with datasets of a specific resolution $s$ corresponding to the \terrain\ size of the training dataset. Working with different dimensions requires training of other networks. In our experiments, we trained two different models and applied them in cascade to increase the initial resolution of a \terrain\ by several orders of magnitude. The $30$-meters resolution model synthesises the structure of large-scale \terrains\ ($90\,\ukm$), whereas the $5$-meters networks are devoted to super-resolution.
Super-resolution and style-mixing can also be combined in order to guide the details towards an exemplar (see \Fig~\ref{fig:sr_result}).

\section{Results and discussion}

\label{subsec:data}

The source used for creating datasets is composed of publicly available raster images of Digital Elevation Models (DEMs). We trained two different \styledem\ at $5$ and $30$ meters resolutions and built the datasets accordingly. 

We used the IGN RGE ALTI database composed of $5 \times 5$ kilometers patches 
with the five-meter precision. We resampled tiles to $1,024 \times 1,024$ using bi-cubic interpolation to adapt to the network requirements. We downloaded $5,600$ DEMs covering France and performed a selection based on their dynamic range so that flat \terrains\ should not be over-represented and over-generated. We created elevation histograms and selected an equal number of representatives for each class. More precisely, every tile was classified into its dynamics rounded to the nearest ten meters. We randomly selected $\approx 20$ \terrains\ to maintain the balance between classes, resulting in $1,760$ patches. 
The second dataset uses elevations from NASA SRTM consisting of $344$ DEMs covering Europe, 
 cut into 3,096 patches at 30 meters precision. We applied the same process to handle the dynamic range and produced $1,900$ patches.

We modified the \stylegan\ to support a 16-bit grayscale precision format for \terrain\ images, which were normalised to unit interval for training purposes.
Because the training of a \stylegan\ is computationally intensive, we relied on a high-performance computing centre.
The generator $\mathcal{G}$, capable of generating \terrains\ with a latent vector $\latentvector$ as input, was trained on 4 Nvidia V100 GPUs with 16 GB of memory, for $35$ hours for the five meters precision and $75$ hours for $30$ meters precision.

\begin{figure*}[h!tb]
  \centering
  \includegraphics[width=\linewidth]{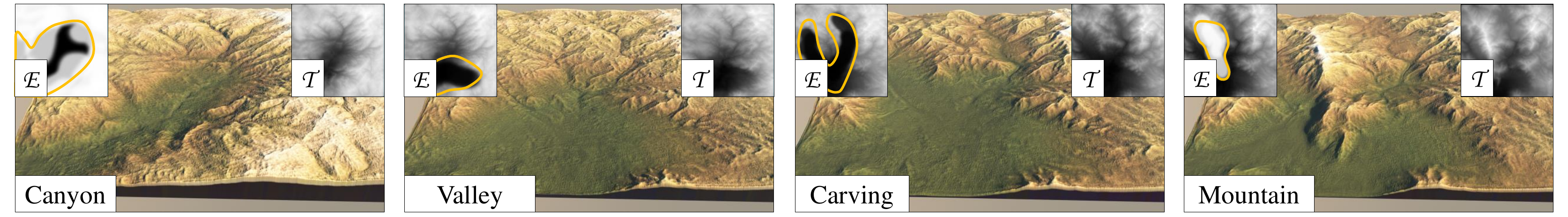}
  \caption{Four steps of a typical $5$ minutes editing session. The user first sketched a low-resolution canyon, and then carved valleys and raised mountains by directly drawing low-resolution elevation level sets in the heightfield. }
  \label{fig:editingsession1}
 \end{figure*}

Once the generator-training process was completed, we generated $20,000$ synthetic \terrains\ to train the encoder using randomly selected latent vectors, divided into $16,000$ images for training and the others for testing. Training was performed on a single NVidia V100 GPU with 16 GB of memory for $12$ hours.

\subsection{Implementation and performance}

The scripts for generating the datasets were implemented in Python, and PyTorch was used for machine learning. \stylegan\ \cite{stylegan2} and the encoder pSp \cite{richardson_encoding_2021} were adapted from the author's implementation. 
The generator runs at $13$ ms on Nvidia GPU hardware (see Table~\ref{table:model_performance}), thus providing interactive feedback to the user. All the terrains throughout the paper have been rendered using Eon-Software Vue with two shading types: realistic (see \Fig~\ref{fig:interpolation}) and cartographic (see \Fig~\ref{fig:sr_result}) when we wanted to emphasise the landforms details.

\begin{table}[t]
 \centering
 \begin{tabular}{|l|r|r|r|r|}\hline
  \tvi Tool & {RTX3090} & {GTX970} & \#$\mathcal{G}$ & \#$\mathcal{I}$ \\ 
				\hline
  \tvi  Generator & 12 ms & 42 ms & 1 & 0 \\ 
  \tvs Encoder & 24 ms & 112 ms & 0 & 1 \\ 
  \tvs \styledem & 37 ms & 163 ms & 1 & 1 \\ 
  \tvs Style mixing & 81 ms & 348 ms & 1& 2 \\\hline
  \tvs Super-resolution $3\times 3$ & 5.2 s & 12.7 s & 25 & 25 \\
  \tvs Super-resolution $6\times 6$ & 16.2 s & 50.6 s & 121 & 121 \\
				\hline
 \end{tabular}
 \caption{Performance for different operations. The last two columns correspond respectively to the number of passes of the generator $\mathcal{G}$ and the encoder $\mathcal{I}$. Note that the super-resolution includes other operations such as blending or retargeting.}
 \label{table:model_performance}
\end{table}

\subsection{Control}

We developed a plugin for the open-source modelling software Blender that grants accessibility to our algorithms to unfamiliar users. It integrates all the functionalities and interactive tools described in \Sec~\ref{sec:authoring_style} and can be used in all the Blender environments: modelling, shaders and render engine. See accompanying video for examples of user interaction.

\begin{figure}[htb]
 \centering
 \includegraphics[width=84mm]{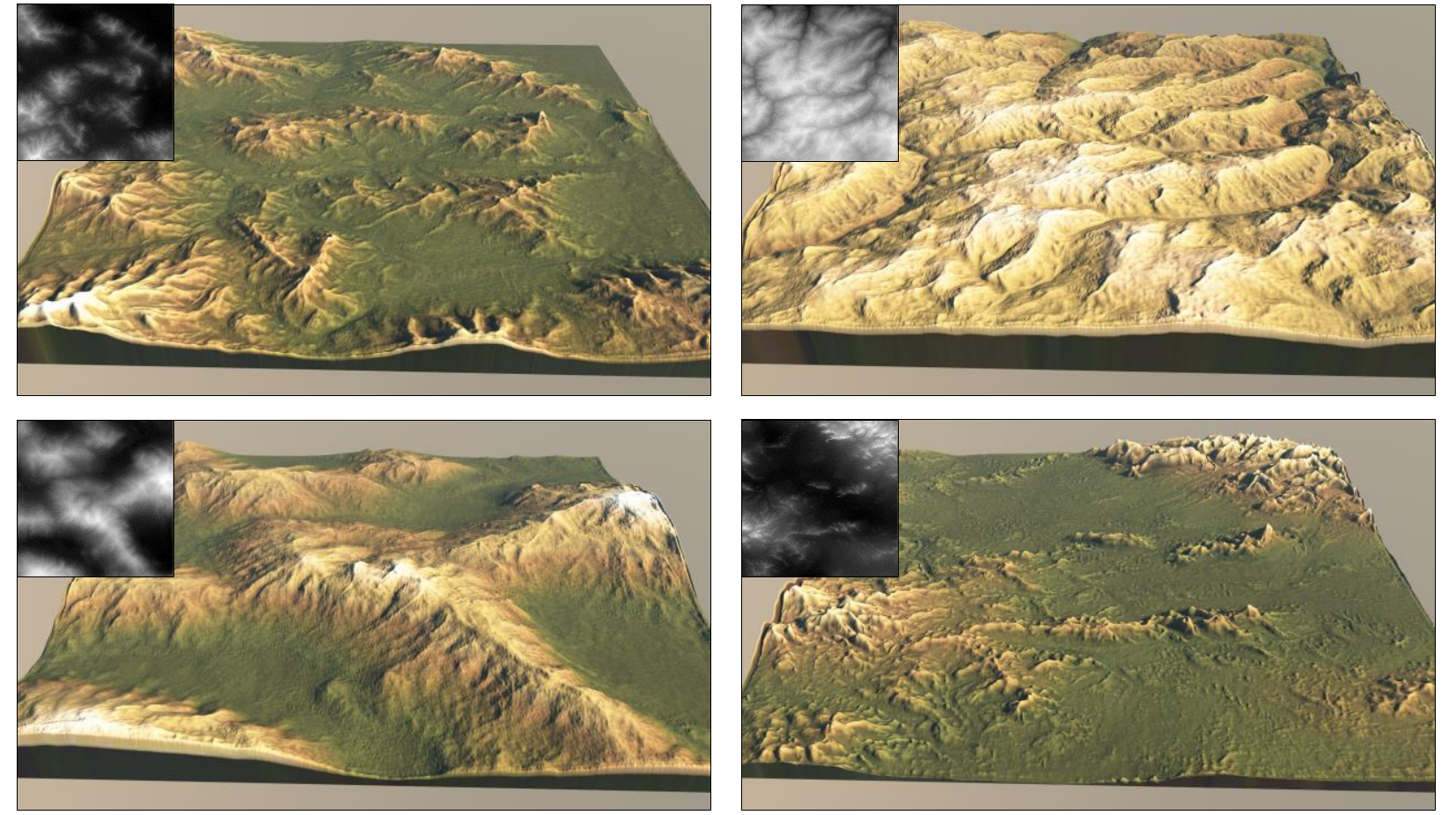}
 \caption{Four terrains obtained by different non-artists users during the qualitative testing process of our model. Only sketching tools have been used during these sessions. }
 \label{fig:editingsessionnonartist}
\end{figure}

To demonstrate the effectiveness of our approach, we conducted a qualitative 
study of our model with $3$ non-artists who were asked to evaluate the modelling tools and 
comment on their effectiveness. 
 \Fig~\ref{fig:editingsessionnonartist} shows examples of terrains produced by untrained users after a $\approx 5$ minutes editing session only, starting from scratch. 
Users \gon reported that they managed to author realistic terrains  \goff following their intent with different styles. \Fig~\ref{fig:editingsession1} shows a typical editing session performed by an experienced designer.

\begin{figure*}[h!tb]
  \centering
  \includegraphics[width=\linewidth]{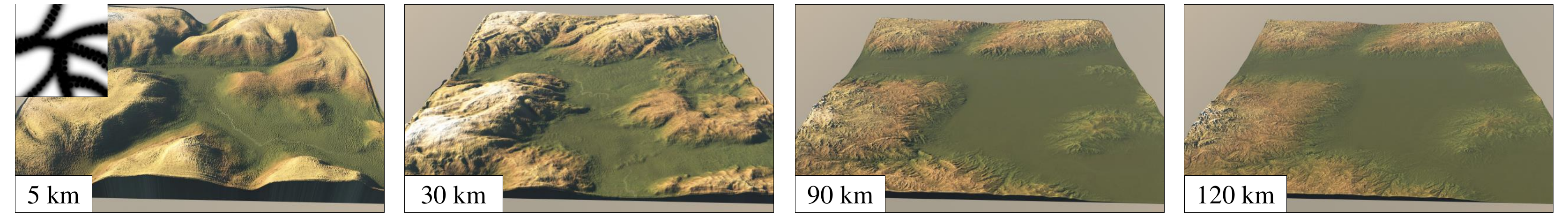}
  \caption{Our framework can adapt to any \terrain\ size by using one tile of the two \styledem\ at $5\,\ukm$ or $30\,\ukm$, or by composing multiple tiles for larger \terrains. }
  \label{fig:sketchdifferentsizes}
 \end{figure*}

 \Fig~\ref{fig:sketchdifferentsizes} shows \gon that our model adapts to different \terrain\ sizes: the same sketch can generate different landforms corresponding to the target scale. \goff

\subsection{Comparison}

\gon Our approach lends itself to terrain authoring and amplification, and \goff compares favourably to state-of-the-art methods with similar goals.
Recently, the Generative Adversarial Terrain amplification (GATA) \cite{Zhao2019} offered a GAN architecture for style embedding and amplification that is considered the state-of-the-art method for terrain amplification. While producing high-quality results by amplifying a low-resolution terrain with a specific input style, it lacks interactive editing tools. The pipeline not only offers style transfer and super-resolution but also proposes multiple authoring tools assembled into a unified framework for building new \terrains\ or modifying existing ones. Moreover, the latent vector representation allows for transfer style at any scale.

\begin{figure}[htb]
  \centering
  \includegraphics[width=84mm]{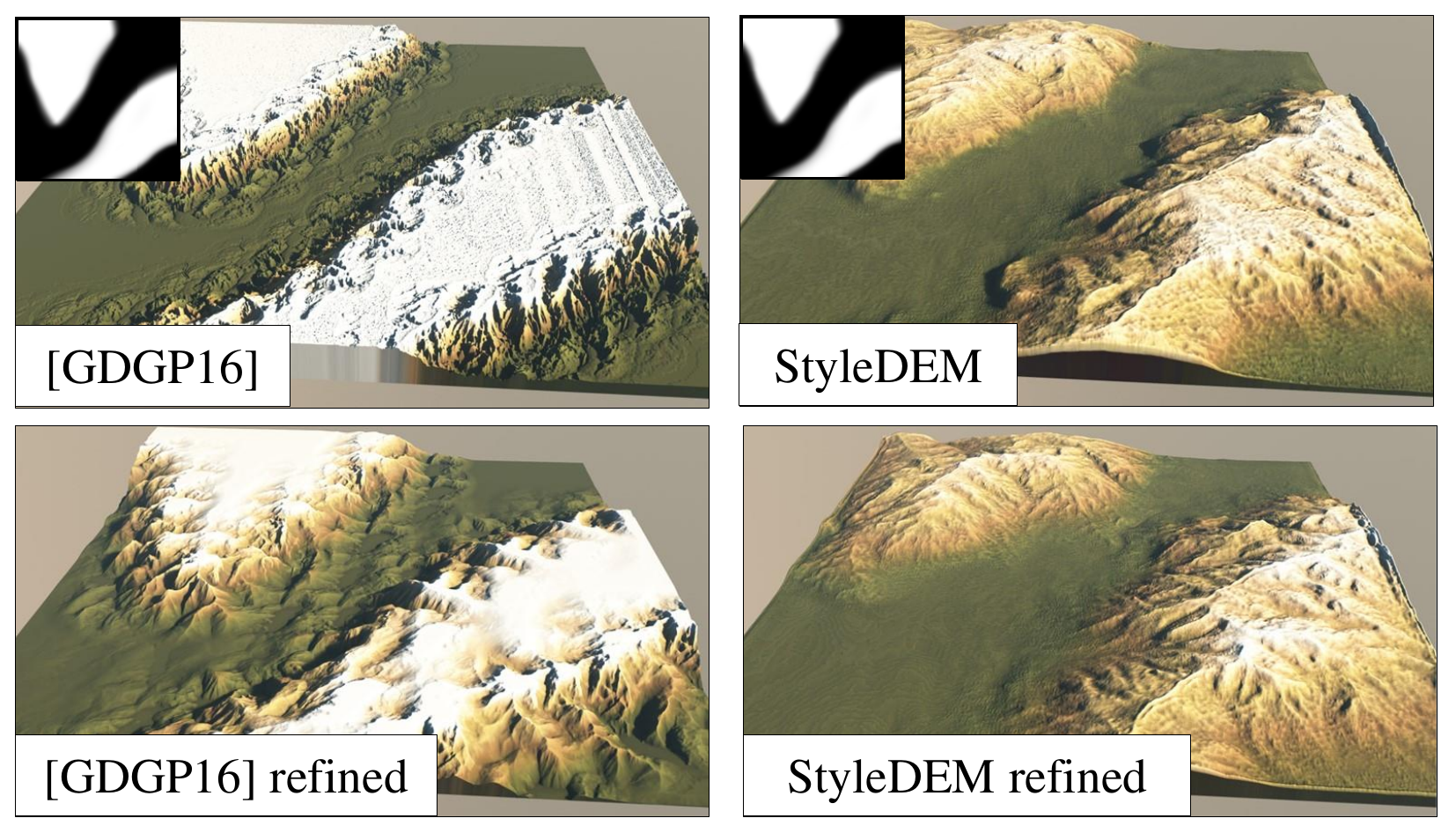}
  \caption{Comparison between sparse modelling \cite{Guerin2016} (left) and StyleDEM (right): our model allows details to be added while keeping global features. The \emph{refined} row corresponds to the sketch going through StyleDEM model twice. }
  \label{fig:comparisonGuerin2016}
 \end{figure}

Amplification and authoring have also been explored by \cite{Guerin2016} and \cite{Argudo2017} by using sparse modelling. However, these approaches do not provide terrain generalisation since patches are selected from exemplars without modification. Combining multiple styles is achievable at the expense of numerous exemplars, which intrinsically limits the performance of the selection in the dictionary. \Fig~\ref{fig:comparisonGuerin2016}, shows that \cite{Guerin2016} introduces repetitions artefacts on planar sketch surfaces, whereas our method manages to generate a palette of consistent landforms.

A recent method \cite{zhang_authoring_2022} implemented a style transfer approach using GANs to encode global and local styles independently. The GAN takes a level set as input and generates a terrain by combining two levels of details. This technique generates convincing variants of different styles. The main limitation comes from the restricted number of styles available, and styles are learned from specific hand-made datasets with evident generalisation limitations. In contrast, our method delivers a large variability by detecting the characteristics of a \terrain\ that was never provided during training. 

Finally, we compare our method to \cite{Guerin2017}, which was, to the best of our knowledge, the first technique proposing a deep-learning framework to \terrain\ generation. Authors adopted a conditional GANs by giving sketches of ridges and rivers and generating high-resolution heightfields. In contrast, we propose a different and more versatile approach for authoring terrains. Style mixing allows fast prototyping using the same input, which would be difficult using the previous method since it requires training another network with a different dataset. 

One crucial facet of our method is the versatility of tools. A single pipeline, with one training, offers extensive possibilities to artists. Table~\ref{table:features_comparison} shows a comparison of the various features available in previous works. We did not include erosion-simulation-based methods that are not relevant in this comparison. 

\begin{table}[!ht]
  \centering
  \begin{tabular}{|l|l|c|c|c|c|c|c|c|c|}
  \hline
  \multicolumn{2}{|c|}{Article} & \rotatebox{90}{Amplification } 
      & \rotatebox{90}{Sketches } 
      & \rotatebox{90}{Interpolation } 
      & \rotatebox{90}{Style-mixing } 
      & \rotatebox{90}{Style brush } 
      & \rotatebox{90}{Curve Constraints } 
      & \rotatebox{90}{Copy-paste } 
      & \rotatebox{90}{Code / Dataset } \\\hline

      \multirow{6}{*}{\rotatebox{90}{Example-based}} & Ours & \textbullet & \textbullet & \textbullet & \textbullet & \textbullet & & \textbullet & \textbullet\ / \textbullet \\ \cline{2-10}
      & \cite{zhang_authoring_2022} & & \textbullet & \textbullet & \textbullet & \textbullet & & & \textbullet\ / \phantom{\textbullet} \\
      & \cite{Zhao2019} & \textbullet & & \textbullet & \textbullet & & & & \textbullet\ / \phantom{\textbullet}  \\
      & \cite{Guerin2017} & \textbullet & \textbullet & &  & &  & \textbullet & / \\
      & \cite{Guerin2016} & \textbullet & \textbullet & & \textbullet & & \textbullet & & \textbullet\ / \phantom{\textbullet}\\
      & \cite{Gain2015} & & \textbullet & & \textbullet & \textbullet & \textbullet & \textbullet & \phantom{\textbullet}\ / - \\\hline

      \multirow{9}{*}{\rotatebox{90}{Procedural}} & \cite{Guerin2022} & \textbullet & \textbullet & & & & \textbullet & \textbullet & \textbullet\ / - \\
      & \cite{gaillard_dendry_2019} & & \textbullet & & & & & & \textbullet\ / - \\
      & \cite{Hnaidi2010} & & & & & & \textbullet & & \phantom{\textbullet}\ / - \\
      & \cite{Genevaux2015} & & & & & & \textbullet & \textbullet & / \\
      & \cite{Argudo2019} & & \textbullet & & \textbullet & & \textbullet & \textbullet & \textbullet\ / \textbullet \\
      & \cite{Gain2009} & & & & & & \textbullet & & \phantom{\textbullet}\ / - \\
      & \cite{Tasse2014} & & & & & & \textbullet & & \textbullet\ / - \\
      & \cite{Carpentier2009} & & \textbullet & & & & & & \phantom{\textbullet}\ / - \\
      & \cite{Genevaux2013} & & \textbullet & & & & \textbullet & & \phantom{\textbullet}\ / - \\\hline

  \end{tabular}
  \caption{Comparison of tools available in our method and other procedural and example-based systems. Dataset release and trained models are only compatible for machine learning methods, '-' means not applicable for this method. 
	}
  \label{table:features_comparison}
\end{table}

\subsection{Validation}

\begin{figure*}[h!tb]
  \centering
  \includegraphics[width=\linewidth]{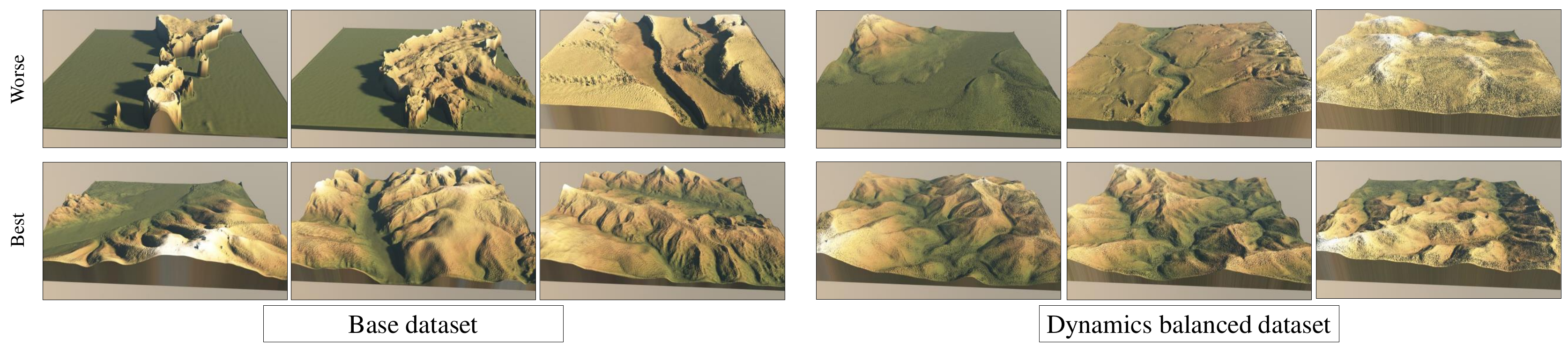}
  \caption{Comparison between unbalanced and balanced dataset. We randomly generated $100$ terrains for each model, and picked the $3$ (a) worst and (b) best-looking examples.}
  \label{fig:failurecases}
 \end{figure*}

We needed to preprocess the dataset to balance the presence of every class. As shown in \Fig~\ref{fig:failurecases}, balancing the dataset significantly reduces the number of failure cases and generates more stable results..

\begin{figure}[htb]
  \centering
  \includegraphics[width=84mm]{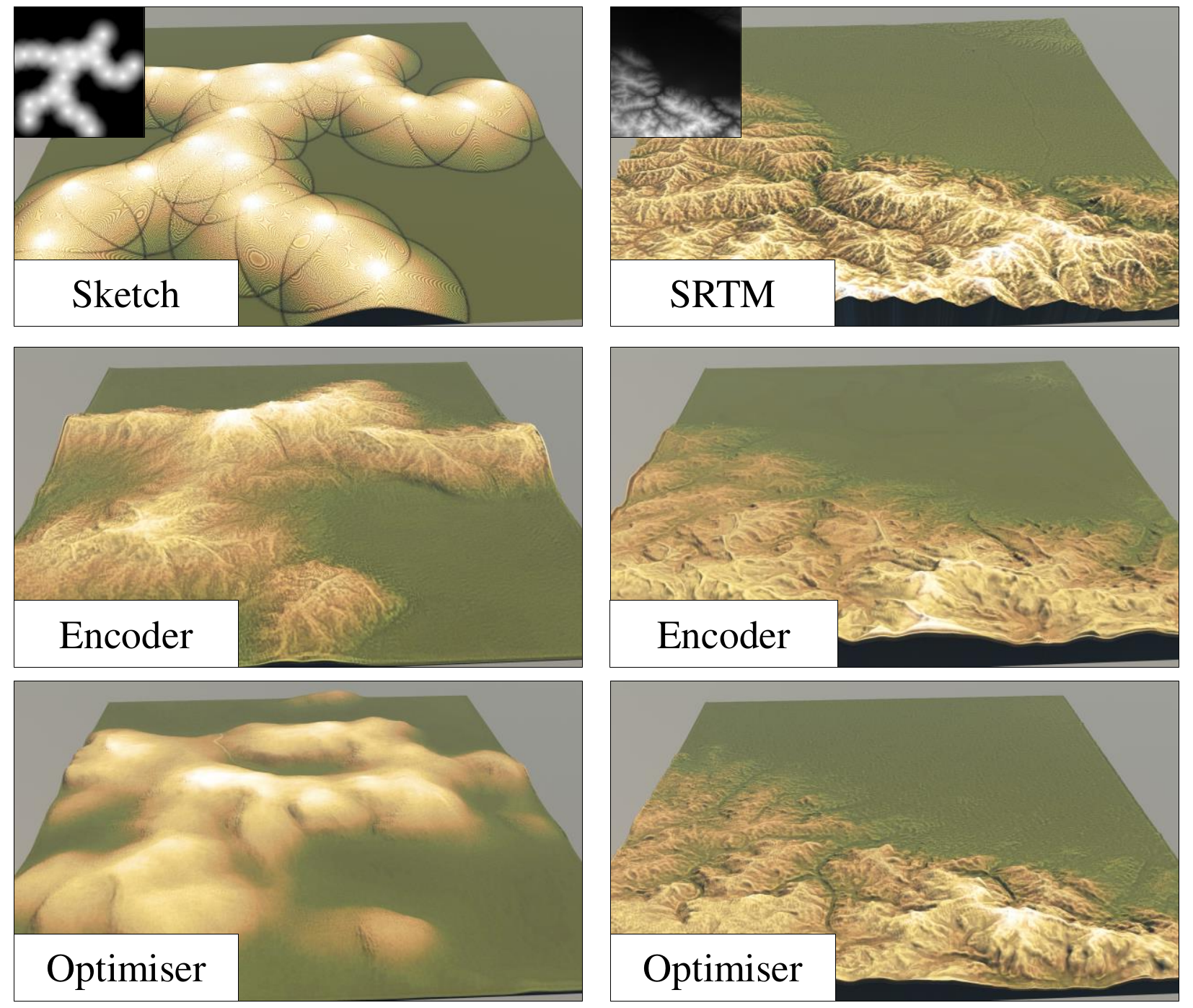}
  \caption{Comparison between encoder and optimiser-based methods for a sketch input (right) and a SRTM DEM (left). The optimiser delivers good results using real \terrains\ and tends to be more accurate in this case. The encoder is a competitor adapted to using sketches and performs better in this configuration by generating features according to the prescribed inputs.  }
  \label{fig:encodervsoptimiser}
\end{figure}

We choose an encoder approach to retrieve the latent vector $\latentvector$ inside the \stylegan\ latent space. 
An alternative method consists in iteratively optimising a random latent vector to match the input data using a standard loss based on the difference between input and the produced \terrain.
Theoretically, this method converts a real \terrain\ into the latent space $\latent$ more precisely and therefore lends itself for style mixing with high-resolution models.
In the case of low-resolution inputs or sketches, the optimisation faithfully reproduces the input terrain without introducing any learned landforms.
By default, the optimisation uses mean squared error on VGG16 features. Adding an L2 loss directly on images yields better results since features such as mountains are  located more precisely. We also found that optimisation is highly dependent on the initialisation.
In contrast, the encoder trained with heightfields allows for a broader range of applications, as exemplified in \Sec~\ref{subsec:handmade_drawing}. 
The optimiser takes $\approx 30\, \us$ to converge to a solution, which is to be compared to  $\approx 37 \,\ums$ using the encoder. Only the latter is compatible with interactive feedback. 
\Fig~\ref{fig:encodervsoptimiser} illustrates a comparison of these approaches.

\begin{figure}[htb]
  \centering
  \includegraphics[width=84mm]{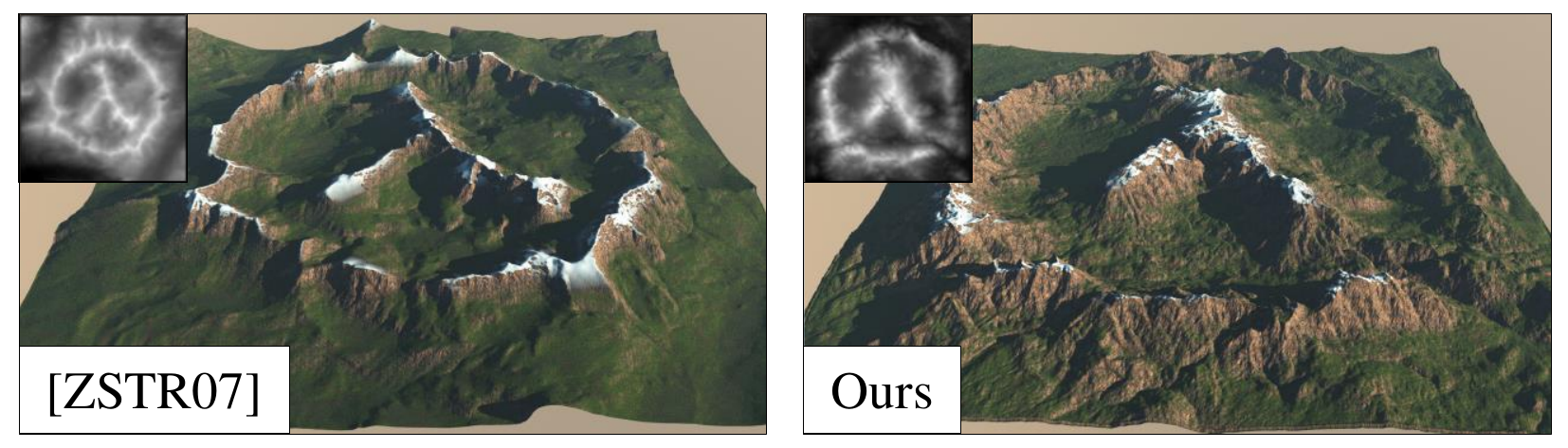}
  \caption{Comparison between texture synthesis \cite{Zhou2007} (left) and \styledem\ (right): our model better preserves landforms at different resolutions. }
  \label{fig:comparisontexturesynthesis}
 \end{figure}

We benefit from the training dataset of \stylegan\ that is only composed of real digital elevation models. 
Visual inspection shows that other data-driven methods generally do not deliver outputs as consistent as erosion simulation approaches.
Moreover, we performed experiments to estimate hydrological and geomorphological coherence by analysing the drainage properties. 
We applied a breaching algorithm to guarantee the drainage discharge over the entire \terrain.
We modified the algorithm to evaluate the volume of bedrock $v_\dem$ removed to enforce drainage consistency.
We compared our technique to other standard methods, and a real digital elevation model in the training dataset, reported in Table~\ref{table:stream_flow}. 
Except \cite{Cordonnier2017} which is a simulation-based method, results demonstrate that $v_\dem$ has perform better than other existing methods. 
This observation is confirmed by the visual inspection: the generator often breaches circular mountain ranges so that water can flow out of them.
\Fig~\ref{fig:comparisontexturesynthesis} shows an example of a sketch featuring an endorheic basin breached by the encoder. 

\subsection{Limitations}
Our approach has several limitations. One comes from the specific size of landforms used to train the generator. A specific encoder and generator need to be trained for every resolution, which requires intensive learning and an increasing amount of exemplar.

Moreover, the encoder does not achieve satisfactory results using small details in the sketch that are not visible in the inversion. After a quantitative study, we found that features smaller than 5\% of the image size, around $50$ pixels for a $1024\times 1024$ terrain, may not be taken into account. Therefore, our method is useful for rapid prototyping, and obtained terrains can be refined with tools provided classical production pipelines. 

\stylegan\ works nicely at generalising objects from a given class. Nevertheless, it fails at inferring classes not presented in the training dataset. The patches used while training contains a single class or style because classes are consistent spatially. 
Thus, the networks cannot generate outputs that combine multiple styles, and the encoder cannot find a corresponding $\latentvector$ for such \terrains. 
While we alleviated this issue by blending outputs of different styles in the elevation domain, this approach remains limited as a post-processing step, which prevents from operating consistently in the latent space. 

Furthermore, every class may not be equally represented in the training dataset, thus leading to quality differences between classes. A tedious manual classification as a processing step would be necessary to conform to this equilibrium.

\begin{table}[h!tb]
  \centering
  \begin{tabular}{|l|r|}
   \hline
         Method & $v_\dem$ \\ \hline
          \cite{Zhou2007} & 440 \\
          \cite{Cordonnier2017} & 1 \\ 
          \cite{Guerin2017} & 178 \\
         SRTM & 57 \\\hline
         Generator $\mathcal{G}$ & 117 \\
         Encoder $\mathcal{I}$ & 97 \\
         \hline
     \end{tabular}
  \caption{The volume of bedrock removed by applying a complete breaching algorithm.}
  \label{table:stream_flow}
 \end{table}


\section{Conclusion}

We introduced \styledem, a versatile deep neural model for terrain authoring that allows designers to perform many editing tasks in the latent space while respecting the overall style of the terrain. From sketch-based authoring to style-transfer, by way of interpolation and super-resolution, the model embeds a description of terrains that inherently encompasses its geomorphological characteristics and guarantees consistency during generation. Experiments and a small-scale user study demonstrate its effectiveness. We provide a complete implementation of the model, along with datasets and a Blender addon that demonstrates the capabilities of this novel representation.


\section*{Acknowledgments}

This work is funded by the project AMPLI ANR-20-CE23-0001, supported by Agence Nationale de la Recherche.
This work was granted access to the HPC resources of IDRIS under the allocation 20XX-AD011013212 made by GENCI.

\bibliography{references}

\end{document}